# A novel Taguchi-based approach for optimizing neural network architectures: application to elastic short fiber composites

Mohammad Hossein Nikzad[a], Mohammad Heidari-Rarani[a,*], Mohsen Mirkhalaf[b,*]

[a] Department of Mechanical Engineering, Faculty of Engineering, University of Isfahan, 81746-73441 Isfahan, Iran

[b] Department of Physics, University of Gothenburg, Origovägen 6B, 41296 Gothenburg, Sweden

**Abstract**

This study presents an innovative application of the Taguchi design of experiment method to optimize the structure of an Artificial Neural Network (ANN) model for the prediction of elastic properties of short fiber reinforced composites. The main goal is to minimize the required computational effort for hyperparameter optimization while enhancing the prediction accuracy. Utilizing a robust design of experiment framework, the structure of an ANN model is optimized. This essentially is the identification of a combination of hyperparameters that yields an optimal predictive accuracy with the fewest algorithmic runs, thereby achieving a significant reduction of the required computational effort. Our findings demonstrate that the Taguchi method not only streamlines the hyperparameter tuning process but also could substantially improve the algorithm's performance. These results underscore the potential of the Taguchi method as a powerful tool for optimizing machine learning algorithms, particularly in scenarios where computational resources are limited. The implications of this study are far-reaching, offering insights for future research in the optimization of different algorithms for improved accuracies and computational efficiencies.

**Keywords:** Artificial neural network; Taguchi design of experiment; Short fiber reinforced composites; Elastic properties.

## 1. Introduction

Over recent decades, classical mechanics and physics models have been developed to predict the complex behavior of materials. Despite their advancements, these models still face the major challenge of high computational costs. To address these limitations, past ten years have seen a significant rise in interest from science and engineering communities in deep learning models, which offer potential solutions for the challenges inherent in traditional modeling and material design approaches (see [1,2] for a comprehensive review). The composite community has also shown a great interest in employing different

*Corresponding authors: mohsen.mirkhalaf@physics.gu.se (M. Mirkhalaf), m.heidarirarani@eng.ui.ac.ir (M. Heidari-Rarani)

Machine Learning (ML) techniques, such as Artificial Neural Networks (ANNs) to address existing challenges in the modelling and design of these materials [3,4].

During the last decade, several studies have developed ANN models for elastic composite materials, see e.g., [5–10]. The studies have used different kinds of neural networks such as Feed-Forward ANNs (FFANNs) and Convolutional Neural Networks (CNNs) with different sources of data to develop efficient and accurate surrogate models for different kinds of composites such as unidirectional, woven, and short fiber composites. More recently, ANN models have also been developed for path-dependent behavior of composite materials, see e.g., [11–16]. These studies use different Recurrent Neural Networks (RNNs), such as Gated Recurrent Units (GRU) and Long Short-Term Memory (LSTM), for the development of their surrogate models. RNNs are well-suited for modeling path-dependent material behavior due to their inherent design and functionality including sequential data handling and memory capability.

Although different ANN architectures have great potential for developing surrogate models and design tools, there are still two major challenges to use these algorithms: (I) ANNs are typically data-hungry [17,18], and (II) optimization of hyperparameters is generally a cumbersome procedure [19,20]. To address the first challenge in the context of material surrogate modelling, recent studies have proposed using different techniques such as (i) using alternative accurate modelling approaches [7], (ii) transfer learning [16,21], and (iii) data augmentation [22]. However, hyperparameter optimization (the second challenge) still imposes difficulties in developing ANN models. To the authors' knowledge, optimizing hyperparameters which involves fine-tuning parameters such as learning rate, batch size, and the number of hidden layers, is usually done (in material science and engineering applications) via either *a trial-and-error* procedure (see e.g., [23,24]) or a *grid search* (see e.g., [16]). These methods are both computationally cumbersome and require a huge amount of computational efforts.

In this study, we are proposing an efficient approach based on the Taguchi design of experiment method [25] for hyperparameter optimization of an ANN model for elastic Short Fiber Reinforced Composites (SFRCs). Taguchi method is typically used for experimental programs to reduce the number of required experiments and thereby, improving the efficiency, and reducing associated costs and material waste [26–28]. The Taguchi method uses orthogonal arrays to systematically choose parameter combinations for experiments in a way that provides a balanced and efficient way to study the effects of multiple factors without requiring a full factorial design. Here, we are proposing to use the Taguchi method for ANN hyperparameter optimization as an alternative to trial-and-error and grid search approaches. We have used this method to develop an ANN model for elastic SFRCs. A comprehensive dataset including micro-structural constitutive and morphological properties and stiffness parameters was taken from [7]. We applied our proposed approach for developing a surrogate ANN model, and our findings show that the method is very effective and considerably reduces the required computational effort for the development of the ANN model. We believe our proposed method for leveraging the Taguchi method for an efficient process of hyperparameter optimization of neural networks can be used not only for material surrogate modelling but also in other fields of engineering and sciences where ANNs have potential applications. The remainder of this paper is structured as follows. Section 2 describes the original dataset taken from [7]. Section 3 describes the Taguchi method. The obtained results and related discussions are given in Section 4. Also, comparisons of the experimental results, taken from the literature, are conducted. Finally, the conclusions of this study and the implications of our findings are given in Section 5.

**2. Original data**

The data points utilized to train and test the ANN algorithm in this study are taken from [7]. In their study, a comprehensive dataset, including 24,540 data points, was generated using a micromechanical two-step homogenization approach. These two homogenization steps are briefly described below.

**Homogenizing unidirectional RVEs:** In the first step, unidirectional (UD) RVEs are generated and spatially discretized, Finite Element (FE) simulations are conducted, and computational homogenization is performed to obtain the homogenized elastic properties. These simulations were conducted using Digimat-FE. A variety of microstructural properties, including different matrix and fiber constitutive properties, fiber volume fractions and different fiber lengths and diameters, were considered in these simulations.

**Orientation Averaging (OA):** Once the homogenized properties of UD RVEs are obtained, orientation averaging is performed in the second step to account for varying fiber orientation distributions within the composite material. For this step, it is required to generate different fiber orientation distributions which is done using Bingham distribution. Figs. 1(a) and 1(b) show a UD RVE and a set of probability distributions of short fibers on a unit sphere, respectively.

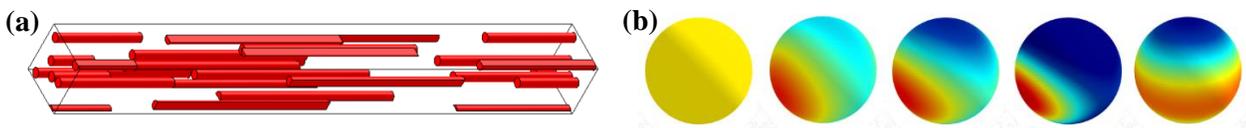

**Fig. 1.** (a) A unidirectional RVE of an SFRC, (b) A representative set of probability distribution functions of short fibers represented on a unit sphere [7].

As mentioned before, a wide variety of microstructural properties were used to conduct simulations for developing the dataset. Table 1 reports the parameter space of the micromechanics simulations. In this table, $E_M$, $\upsilon_M$, $E_F$, $\upsilon_F$ are matrix and fiber elastic modulus and Poisson's ratio; $d_f$, $\lambda_f$ are fiber diameter and aspect ratio; $\phi$ is the fiber volume fraction; $a_{11}$, $a_{22}$ are diagonal components of orientation tensor; $\gamma_1$, $\gamma_2$, and $\gamma_3$ refer to rotations angles of diagonal orientation tensor around three main axes. For more detailed information, please see [7].

Table 1 Microstructural parameters space for the micromechanical simulations [7]

| Parameter | Minimum | Maximum |
| --- | --- | --- |
| $E_M$ (MPa) | 500 | 20000 |
| $\upsilon_M$ (−) | 0.30 | 0.49 |
| $E_F$ (MPa) | 10000 | 100000 |
| $\upsilon_F$ (−) | 0.2 | 0.4 |
| $d_f$ (μm) | 4 | 20 |
| $\lambda_f$ (−) | 2 | 100 |
| $\phi$ (−) | 0.05 | 0.3 |
| $a_{11}$ (−) | 0.33 | 1.0 |
| $a_{22}$ (−) | $1 - 2a_{11}$ | $a_{11}$ |
| $\gamma_1$ (rad) | 0 | Ω |
| $\gamma_2$ (rad) | 0 | Ω |
| $\gamma_3$ (rad) | 0 | Ω |

## 3. Taguchi-based ANN model

Design of Experiment (DOE) is a statistical technique that evaluates how independent variables influence dependent variables, also known as responses [29,30]. DOE is typically employed in empirical studies to obtain optimal results using the fewest possible experiments. The well-known methods within DOE are the Response Surface Methodology [31–33], Mixture Design of Experiment [34], and Taguchi Design of Experiment [35,36]. Notably, the Taguchi method is acclaimed for its ability to finely tune independent variables to improve outcomes efficiently. In this research, it is proposed to use the Taguchi method to enhance the structure of an ANN model, aiming to optimize the prediction accuracy while minimizing the computational cost and the number of required runs. The hyperparameters of the ANN model are considered independent variables to be optimized using the Taguchi method. Table 2 gives the levels considered for the hyperparameters of the ANN algorithm. In this Table, HL, NN, ACT, OPT, and

LR refer to the number of hidden layers, the number of neurons in each hidden layer, activation function, optimizer, and learning rate, respectively

Table 2 Different levels of the ANN hyperparameters

| Factors | Levels | | |
|---|---|---|---|
| | One | Two | Three |
| HL | 1 | 2 | 3 |
| NN | 10 | 20 | 30 |
| ACT | relu | elu | selu |
| OPT | Adam | Adamax | RMSprop |
| LR | 0.001 | 0.01 | 0.1 |

In this study, we considered one, two, and three as the choices of the number of hidden layers. One hidden layer is suitable for simpler problems where a single transformation can effectively map inputs to outputs. It reduces the risk of overfitting and computational cost. Two hidden layers introduce an additional level of complexity, allowing the network to learn intermediate representations, which can improve performance for moderately complex tasks. Three hidden layers further increase the network's capacity to model complex functions. It is beneficial for capturing deep, hierarchical patterns in different datasets, though it requires careful regularization to prevent overfitting. Moreover, 10, 20, and 30 are considered for the number of neurons in each hidden layer. 10 neurons provide a small, efficient network suitable for less complex problems. It helps in quick training and reduces the risk of overfitting. The second choice, 20 neurons, balances between learning capacity and computational efficiency. This level is often sufficient for a wide range of moderately complex tasks. The final level, 30 neurons, offers an increased capacity to capture more detailed patterns in the data, which is useful for complex problems but may require more computational resources and careful tuning to avoid overfitting. In addition, relu, elu, and selu are considered as the levels of the activation function. Relu is popular due to its simplicity and effectiveness in mitigating the vanishing gradient problem. It speeds up training but can suffer from dead neurons. Elu helps with faster learning and more robust performance by maintaining mean activations closer to zero and mitigating the dying relu problem. Selu normalizes outputs automatically and helps maintain a self-normalizing network, which can lead to better performance and faster convergence in deeper networks. Also, Adam, Adamax, and RMSprop are considered for the levels of the optimizer. Adam combines the advantages of two other extensions of stochastic gradient descent, AdaGrad, and RMSprop, making it well-suited for problems with sparse gradients. Adamax is a variant of Adam based on the infinity norm, which can sometimes offer better performance, particularly when dealing with certain types of data distributions. RMSprop is an adaptive learning rate method designed to work well in non-stationary environments. It adjusts the learning rate based on the moving average of recent gradient magnitudes, making it suitable for complex, noisy data. Finally, three values of 0.001, 0.01, and 0.1 are considered for the levels of the learning rate. A learning rate of 0.001 is a small learning rate that ensures stable convergence, useful for fine-tuning the network. It reduces the risk of overshooting the minimum. A learning rate of 0.01 is a moderate learning rate that offers a balance between speed and stability, often leading to faster convergence without compromising accuracy. A learning rate of 0.1 is a higher learning rate that can accelerate the convergence process, especially in the early stages of training. However, it requires careful monitoring to prevent overshooting and divergence.

According to Table 2, a total number of $3^5 = 243$ cases should be examined for the ANN algorithm to find the best hyperparameters. However, just $3^3 = 27$ cases should be evaluated by the implementation of the Taguchi method to find an appropriate structure for the ANN algorithm which significantly reduces the computational costs. For this purpose, an L27 Taguchi orthogonal array is designed for the optimization of the ANN structure using Minitab software (see Table 3).

Table 3 L27 Taguchi orthogonal array

| Run order | HL | NN | ACT | OPT | LR |
|---|---|---|---|---|---|
| 1 | 1 | 10 | relu | Adam | 0.001 |
| 2 | 1 | 10 | relu | Adam | 0.010 |
| 3 | 1 | 10 | relu | Adam | 0.100 |
| 4 | 1 | 20 | elu | RMSprop | 0.001 |
| 5 | 1 | 20 | elu | RMSprop | 0.010 |
| 6 | 1 | 20 | elu | RMSprop | 0.100 |
| 7 | 1 | 30 | selu | Adamax | 0.001 |
| 8 | 1 | 30 | selu | Adamax | 0.010 |
| 9 | 1 | 30 | selu | Adamax | 0.100 |
| 10 | 2 | 10 | elu | Adamax | 0.001 |
| 11 | 2 | 10 | elu | Adamax | 0.010 |
| 12 | 2 | 10 | elu | Adamax | 0.100 |
| 13 | 2 | 20 | selu | Adam | 0.001 |
| 14 | 2 | 20 | selu | Adam | 0.010 |
| 15 | 2 | 20 | selu | Adam | 0.100 |
| 16 | 2 | 30 | relu | RMSprop | 0.001 |
| 17 | 2 | 30 | relu | RMSprop | 0.010 |
| 18 | 2 | 30 | relu | RMSprop | 0.100 |
| 19 | 3 | 10 | selu | RMSprop | 0.001 |
| 20 | 3 | 10 | selu | RMSprop | 0.010 |
| 21 | 3 | 10 | selu | RMSprop | 0.100 |
| 22 | 3 | 20 | relu | Adamax | 0.001 |
| 23 | 3 | 20 | relu | Adamax | 0.010 |
| 24 | 3 | 20 | relu | Adamax | 0.100 |
| 25 | 3 | 30 | elu | Adam | 0.001 |
| 26 | 3 | 30 | elu | Adam | 0.010 |
| 27 | 3 | 30 | elu | Adam | 0.100 |

## 4. Results and discussion

Before optimization of the hyperparameters of ANN algorithms using the Taguchi method, all inputs including parameters of Table 1, and outputs including the components of the stiffness tensor of anisotropic SFRCs are normalized utilizing the Scikit-learn library in Python. Then, 80%, 15%, and 5% of data points are randomly allocated for train, validation, and test sets, respectively. The R-squared value of the train set is considered the criterion for optimizing the hyperparameters of the ANN algorithm. Table 4 reports the R-squared values of all 27 states of the L27 Taguchi orthogonal array.

Table 4 R-squared values of the L27 Taguchi orthogonal array

| Run order | $R^2$ of train, % |
|---|---|
| 1 | 68.295 |
| 2 | 67.398 |
| 3 | 57.679 |
| 4 | 83.541 |
| 5 | 78.183 |
| 6 | 46.093 |
| 7 | 71.553 |
| 8 | 67.558 |
| 9 | 31.665 |
| 10 | 79.734 |
| 11 | 81.528 |
| 12 | 79.261 |
| 13 | 93.682 |
| 14 | 89.224 |
| 15 | 74.192 |
| 16 | 92.747 |
| 17 | 89.989 |
| 18 | 36.973 |
| 19 | 80.197 |
| 20 | 78.136 |
| 21 | 29.542 |
| 22 | 92.436 |
| 23 | 92.835 |
| 24 | 85.405 |
| 25 | 98.680 |
| 26 | 92.617 |
| 27 | 79.556 |

Notably, the early stopping criterion with patience = 200 is considered to avoid overfitting the ANN algorithm. Early stopping is a technique employed to terminate the training of neural networks at an optimal point, ensuring they perform effectively on both the training and validation data points. The "patience" parameter determines how many epochs the training will continue without an improvement in the validation performance before the training is stopped. This prevents the model from stopping too early, which could lead to an underfit model, and also prevents the model from training for too long, which could lead to an overfit model. During neural network training, the model's performance is continuously assessed by monitoring the loss on a separate validation set. Typically, as training progresses, the loss on the training set decreases, indicating effective learning. However, if the model starts to overly memorize the training data, it may begin to perform poorly on the validation set. The early stopping mechanism detects this by observing if the validation loss starts to increase, indicating a decline in the model's generalization ability. Specifically, with a patience parameter set to 200 epochs, the training process is allowed to continue for 200 additional epochs after the initial increase in validation loss. If the validation loss continues to rise during these 200 epochs, it confirms that the model is overfitting, and training is subsequently halted. This approach helps to ensure the model maintains its ability to generalize to new data, optimizing its performance on datasets beyond the training set. Fig. 2 illustrates the Taguchi-proposed structure for the ANN algorithm.

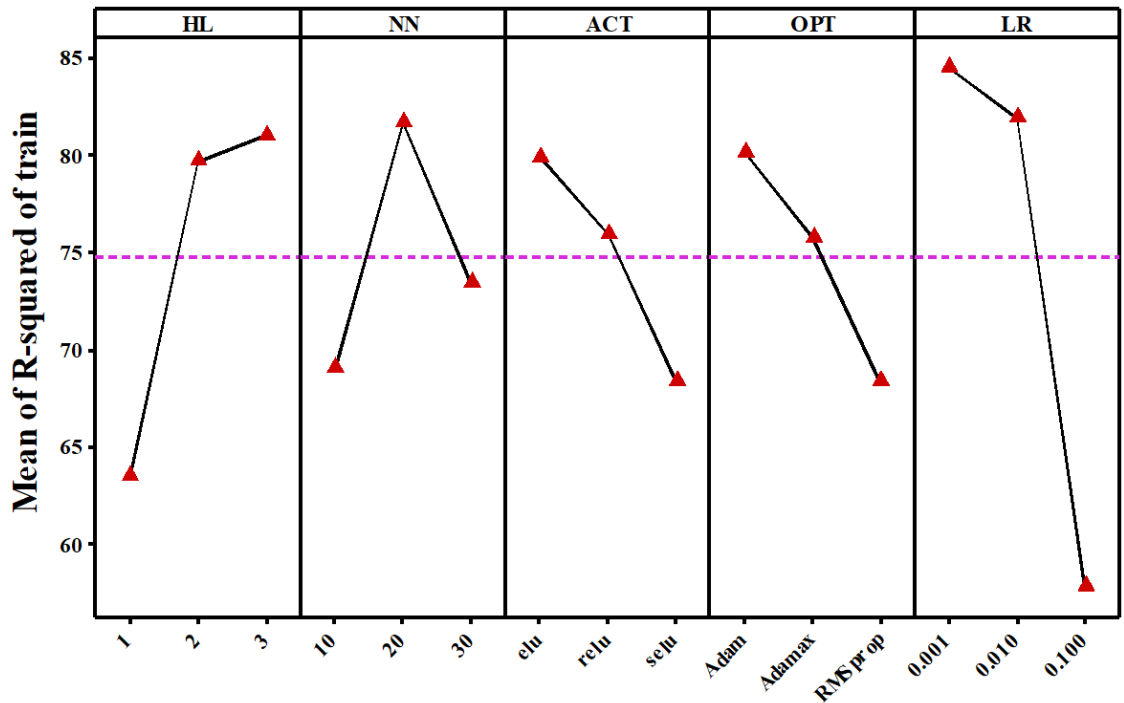

**Fig. 2.** The Taguchi-proposed ANN structure for anticipating the stiffness matrix components of SFRCs.

According to Fig. 2, it is expected that the ANN algorithm has the highest $R^2$ value for the train set with three hidden layers, 20 neurons in each hidden layer, the "elu" activation function, "Adam" optimizer, and learning rate = 0.001. The Taguchi-proposed ANN architecture was created, and the training/validation process was conducted. Fig. 3 shows the loss convergence plot of the ANN model.

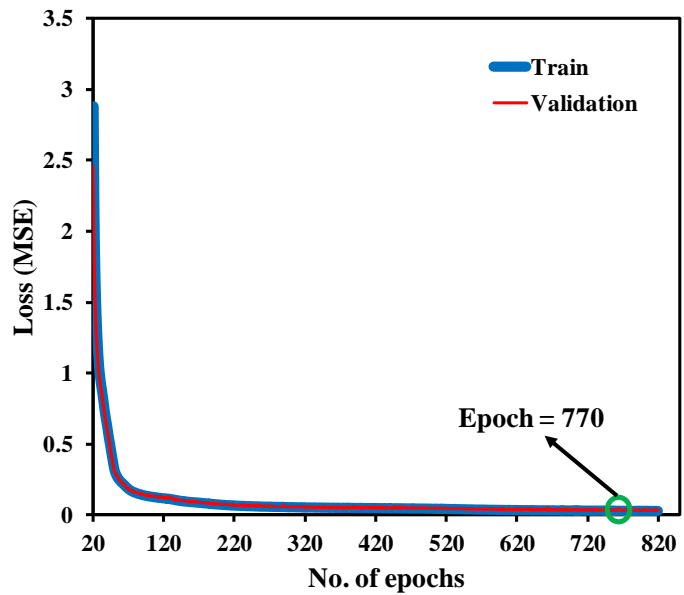

**Fig. 3.** Loss-convergence plot of the Taguchi-proposed ANN structure.

It can be found from this figure that the training of this ANN structure is stopped at the epoch = 770. For evaluating the predictive performance of the Taguchi-proposed ANN structure, four evaluation matrices of R-squared ($R^2$), Mean Absolute Error (MAE), Mean Squared Error (MSE), and Root Mean Squared error (RMSE) are considered. The values of these four evaluation matrices are calculated as follows:

$$R^2 = \left(1 - \frac{\sum_{i=1}^{n}(y_i - \hat{y}_i)^2}{\sum_{i=1}^{n}(y_i - \bar{y})^2}\right) \times 100, \tag{1}$$

$$\text{MAE} = \frac{1}{n}\sum_{i=1}^{n}|y_i - \hat{y}_i|, \tag{2}$$

$$\text{MSE} = \frac{1}{n}\sum_{i=1}^{n}(y_i - \hat{y}_i)^2, \tag{3}$$

$$\text{RMSE} = \sqrt{\frac{1}{n}\sum_{i=1}^{n}(y_i - \hat{y}_i)^2}. \tag{4}$$

In Equations (1)-(4), $y_i$, $\hat{y}_i$, $n$, and $\bar{y}$ are the actual response, predicted response values, the number of observations, and the mean of the actual values, respectively. Table 5 reports the $R^2$, MAE, MSE, and RMSE values of the train, validation, and test sets obtained from the Taguchi-proposed ANN structures for simultaneous prediction of all components of the stiffness matrix of SFRCs.

**Table 5** Evaluation matrices of the Taguchi-proposed ANN structure for the simultaneous prediction of the components of SFRCs stiffness matrix

| Set | $R^2$, % | MAE | MSE | RMSE |
| --- | --- | --- | --- | --- |
| Train | 98.543 | 0.069 | 0.014 | 0.12 |
| Validation | 97.509 | 0.076 | 0.024 | 0.157 |
| Test | 97.708 | 0.076 | 0.024 | 0.157 |

According to the evaluation matrices reported in Table 5 for the test set, it can be found that the Taguchi-proposed ANN structure has an appropriate predictive performance for unseen data points. In the following, the regression plots of all components of the stiffness tensor obtained from the Taguchi-based ANN structure are shown for the train and test sets in Figs. 4 and 5, respectively. Furthermore, the $R^2$, MAE, MSE, and RMSE values of each component for train and test sets are reported in Table 6.

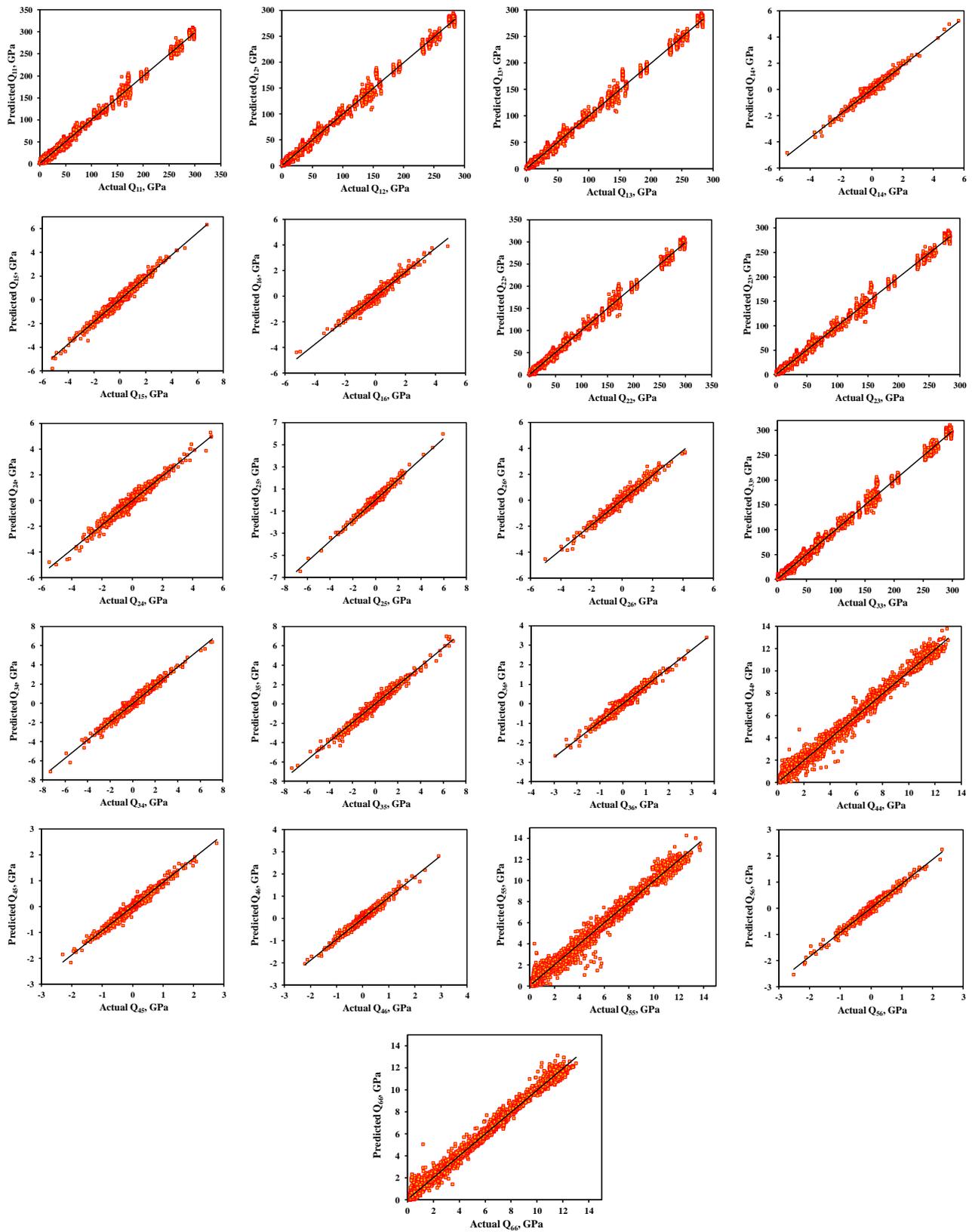

**Fig. 4.** Regression plots for all of the independent components of the stiffness tensor for the training set.

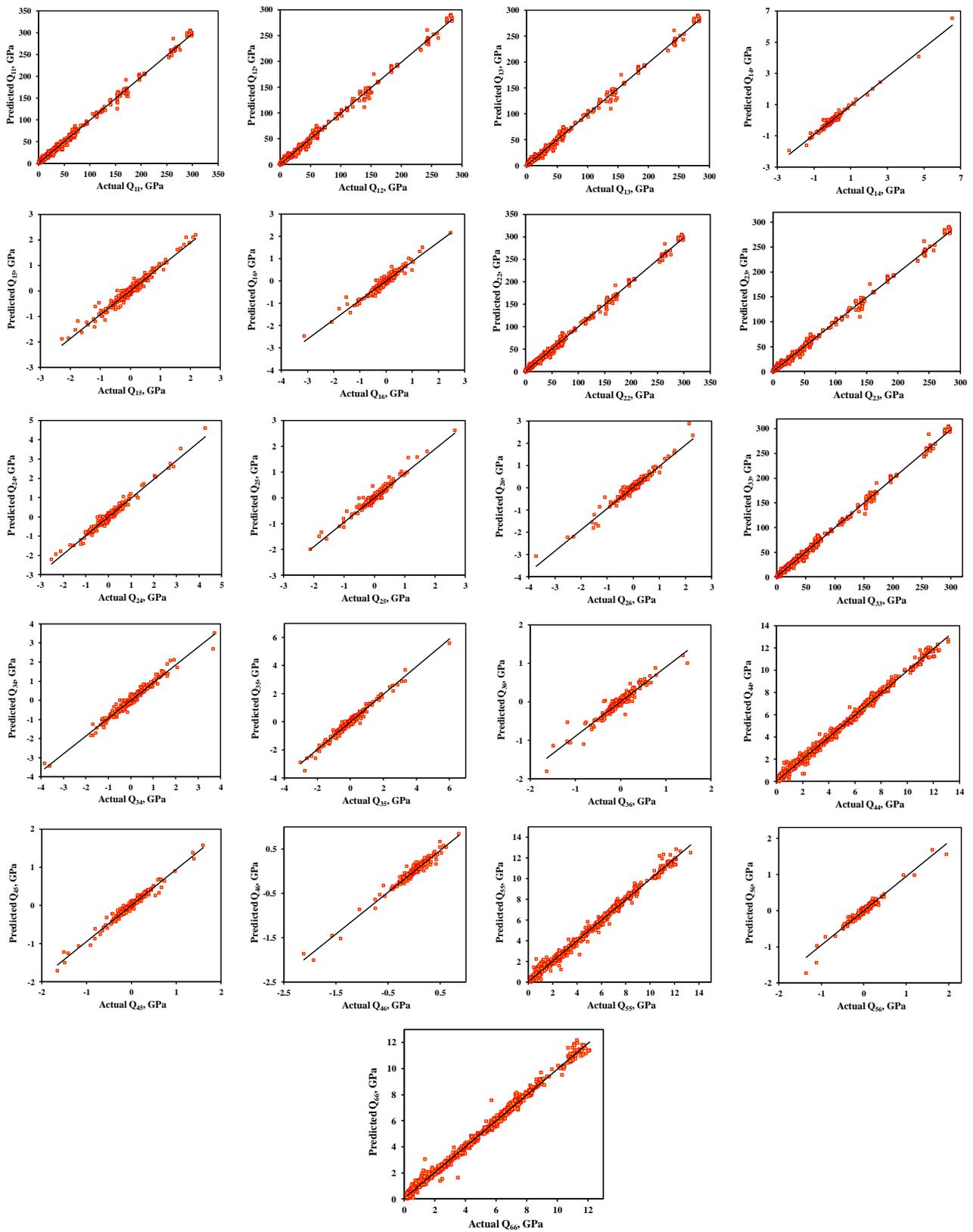

**Fig. 5.** Regression plots for all of the components of the stiffness tensor for the test set.

Table 6 $R^2$, MAE, MSE, and RMSE values for both train and test sets for each stiffness tensor component.

| Component | Train | | | | Test | | | |
|---|---|---|---|---|---|---|---|---|
| | $R^2$, % | MAE | MSE | RMSE | $R^2$, % | MAE | MSE | RMSE |
| $Q_{11}$ | 99.8 | 0.035 | 0.002 | 0.049 | 99.8 | 0.036 | 0.002 | 0.05 |
| $Q_{12}$ | 99.8 | 0.034 | 0.002 | 0.048 | 99.8 | 0.036 | 0.002 | 0.05 |
| $Q_{13}$ | 99.8 | 0.034 | 0.002 | 0.047 | 99.8 | 0.036 | 0.002 | 0.05 |
| $Q_{14}$ | 97.2 | 0.097 | 0.028 | 0.168 | 97.6 | 0.112 | 0.051 | 0.226 |
| $Q_{15}$ | 97.7 | 0.093 | 0.023 | 0.153 | 94 | 0.111 | 0.048 | 0.218 |
| $Q_{16}$ | 97.8 | 0.095 | 0.022 | 0.148 | 94.9 | 0.109 | 0.043 | 0.208 |
| $Q_{22}$ | 99.8 | 0.033 | 0.002 | 0.046 | 99.8 | 0.034 | 0.002 | 0.048 |
| $Q_{23}$ | 99.8 | 0.034 | 0.002 | 0.048 | 99.8 | 0.036 | 0.002 | 0.05 |
| $Q_{24}$ | 97.5 | 0.096 | 0.025 | 0.157 | 96.1 | 0.113 | 0.044 | 0.21 |
| $Q_{25}$ | 97.5 | 0.095 | 0.025 | 0.158 | 95.1 | 0.108 | 0.041 | 0.203 |
| $Q_{26}$ | 97.9 | 0.091 | 0.021 | 0.146 | 97 | 0.099 | 0.031 | 0.176 |
| $Q_{33}$ | 99.7 | 0.037 | 0.003 | 0.051 | 99.7 | 0.039 | 0.003 | 0.054 |
| $Q_{34}$ | 98 | 0.088 | 0.02 | 0.141 | 96.7 | 0.1 | 0.031 | 0.175 |
| $Q_{35}$ | 98.1 | 0.086 | 0.019 | 0.139 | 97.5 | 0.099 | 0.027 | 0.165 |
| $Q_{36}$ | 97.2 | 0.102 | 0.028 | 0.168 | 94.9 | 0.116 | 0.041 | 0.201 |
| $Q_{44}$ | 99.7 | 0.042 | 0.003 | 0.056 | 99.7 | 0.04 | 0.003 | 0.054 |
| $Q_{45}$ | 97.8 | 0.088 | 0.022 | 0.15 | 97.3 | 0.098 | 0.029 | 0.17 |
| $Q_{46}$ | 97.5 | 0.098 | 0.025 | 0.158 | 97 | 0.104 | 0.035 | 0.186 |
| $Q_{55}$ | 99.7 | 0.042 | 0.003 | 0.057 | 99.7 | 0.041 | 0.003 | 0.056 |
| $Q_{56}$ | 97.7 | 0.092 | 0.023 | 0.15 | 96.3 | 0.102 | 0.037 | 0.192 |
| $Q_{66}$ | 99.6 | 0.047 | 0.004 | 0.063 | 99.6 | 0.047 | 0.004 | 0.063 |

According to Table 6, all components of the stiffness tensor of SFRCs are predicted well via the obtained ANN structure. In the following, to assess the predictive performance of the Taguchi-based ANN algorithm on the experimental data points, five experimental data points are collected from [37,38] with the mechanical properties reported in Table 7.

Table 7 Mechanical properties of experimental data points (taken from [38,39])

| Parameter | $d_f$ (μm) | $l_f$ (μm) | ϕ | $E_M$ (GPa) | $\upsilon_M$ | $E_F$ (GPa) | $\upsilon_F$ | $a_{11}$ | $a_{22}$ | $a_{33}$ |
|---|---|---|---|---|---|---|---|---|---|---|
| **Polypropylene/flax 1** [37] | 16 | 1200 | 0.13 | 1.6 | 0.4 | 69 | 0.15 | 0.333 | 0.333 | 0.333 |
| **Polypropylene/flax 2** [37] | 16 | 1200 | 0.21 | 1.6 | 0.4 | 69 | 0.15 | 0.333 | 0.333 | 0.333 |
| **Polypropylene/flax 3** [37] | 16 | 1200 | 0.29 | 1.6 | 0.4 | 69 | 0.15 | 0.333 | 0.333 | 0.333 |
| **Polyamide/15wt.% glass** [38] | 13.5 | 430 | 0.064 | 2.8 | 0.4 | 70 | 0.2 | 0.507 | 0.473 | 0.02 |
| **Polyamide/30wt.% glass** [38] | 12.6 | 366 | 0.152 | 2.8 | 0.4 | 70 | 0.2 | 0.604 | 0.354 | 0.042 |

These five experimental data points are modelled via the Taguchi-based ANN model and the predicted and actual $E_{11}$ and $E_{22}$ values are reported in Fig. 6. As can be observed in this figure, the error between the experimental and ANN-predicted values for Polypropylene/flax 1, Polypropylene/flax 2, Polypropylene/flax 3, Polyamide/15wt.% glass, and Polyamide/30wt.% glass are 5.778%, 4.886%, 6.558%, 8.529%, and 4.686%, respectively. Moreover, the differences between the actual and predicted $E_{22}$ values for Polyamide/15wt.% glass and Polyamide/30wt.% glass are 8.339% and 4.358%, respectively. These predictions show the reliability of the ANN model predictions. As a result, it can be said the proposed Taguchi-based approach for ANN architecture and hyperparameter optimization has great potentials for developing accurate data-driven models with a remarkably reduced computational effort.

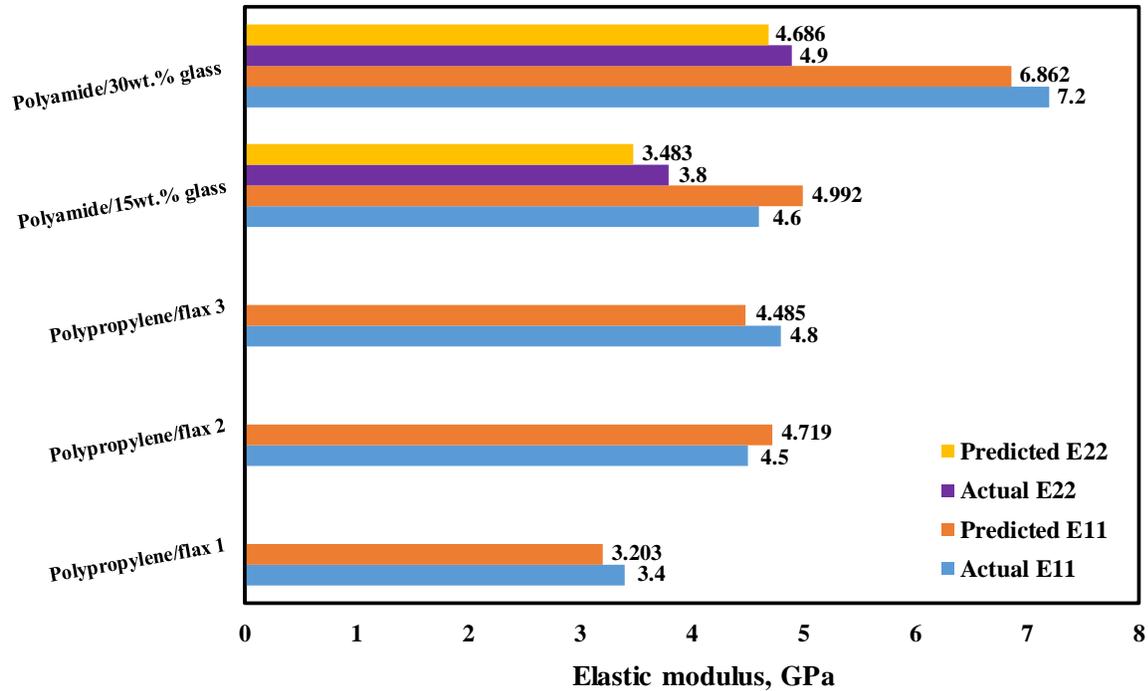

**Fig. 6.** predicted and actual $E_{11}$ and $E_{22}$ values for experimental data points

## 5. Conclusion

This study successfully demonstrated the effectiveness of the Taguchi method in optimizing the structure of artificial neural networks for the prediction of elastic properties in short fiber-reinforced composites. By systematically exploring the hyperparameter space using a robust experimental design framework, the proposed approach was able to identify an optimal ANN configuration that maximizes predictive accuracy while minimizing computational efforts. Instead of a cumbersome trial-and-error approach or a grid search with comprehensive full-factorial analysis, the proposed approach only requires a limited number of tries, and hence, dramatically reduces the required time and efforts for the development of an ANN model. The good error metrics and comparisons to experimental results show the reliability of the proposed method. We believe the Taguchi-based method represents an important advancement in the field of composite materials modeling. In addition to the specific application analyzed in this work, the findings of this study highlight the broader potential of the proposed approach for ANN optimizations for a variety of tasks not only in data-driven modelling and design of materials but also in other scientific and industrial applications.

**Acknowledgments**

Mohsen Mirkhalaf gratefully acknowledges financial support from the Swedish Research Council (VR grant: 2019-04715).